# A Robot Expressing Emotions Through Gestures: Everyone Outside of Italy Would Understand this?

Ilaria Consoli

University of Turin, ilaria.consoli@edu.unito.it

Claudio Mattutino

Dept. of Computer Science, University of Turin, claudio.mattutino@unito.it

Cristina Gena

Dept. of Computer Science, University of Turin, cristina.gena@unito.it

In the context of our research activities on affective computing and human-robot interaction we are working on both the recognition of human's emotions and the expression of emotions by robots. In our vision, robots will be increasingly present in schools, factories, and homes, and their empathetic behavior may foster their acceptance. In particular, in one of our research, we sought to replicate gestures associated with specific emotions on a social robot, NAO. Our focus was on Ekman's six primary emotions, along with five emotions selected from Plutchik's wheel of emotions. In our opinion the cultural component linked to the expression of emotions through gestures certainly influenced both us and the participants. Thus, we would like to investigate the influence of our culture in the gestural expression of emotion.

## 1 PURPOSE

Human communication can be categorized into verbal, encompassing speech, and non-verbal, commonly referred to as 'body language.' Non-verbal communication is deemed the most crucial facet of human interaction, considering that 38% of a person's messages are conveyed through para-verbal communication (tone, volume, rhythm...), 55% through body language, and only 7% through speech [13]. The field of study dedicated to understanding body language is known as kinesics, providing insights into the meanings behind facial expressions, gaze, poses, and gestures.

The face, being an exceptionally expressive part of our body, universally represents a broad spectrum of emotions such as sadness, anger, fear, happiness, surprise, and disgust, which are emotions also known as the Ekman's six primary emotions [3]. For instance, a pronounced smile unmistakably signals happiness, while wide eyes and raised eyebrows convey surprise. Additionally, the face can convey a blend of emotions, as seen in expressions like fright, which simultaneously express fear and surprise [3].

Body posture and gestures also offer valuable clues about an individual's emotional state. Examining specific aspects like arm positioning can reveal various messages. For instance, crossed arms may indicate a closed attitude, while a slouched posture may signify sadness. Basic gestures, like smiling when happy or frowning when angry, are universally understood across cultures. Gestures can be categorized as intrinsic or extrinsic [10]. Intrinsic gestures, like nodding in agreement, are innate and have evolved as stable behavioral patterns through natural selection. Extrinsic gestures, such as turning sideways as a sign of rejection, are learned during early childhood.

Certain emotions are more effectively expressed through facial expressions, while others find better communication through body movements.

In the context of our research activities on affective computing and human-robot interaction we are working on both the recognition of human's emotions [8] and the expression of emotions by robots [7]. In our vision, robots will be increasingly present in schools [6], factories [1], and homes [12], and their empathetic behavior may foster their acceptance [9].

In particular, in one of our research, we sought to replicate gestures associated with specific emotions on a social robot, NAO. Our focus was on Ekman's six primary emotions [3], along with five emotions selected from Plutchik's wheel of emotions [14].

Initially, eleven emotions were considered, with six being Ekman's main emotions:
- Disgust
- Happiness
- Fear
- Anger
- Surprise
- Sadness

Additionally, five emotions were chosen from Plutchik's wheel of emotions:
- Love
- Interest
- Disapproval
- Boredom
- Thoughtfulness

Then the total number of emotions represented was consistently more than eleven due to some emotions being depicted multiple times, see Figure 1. The sensory aspect most engaged in the experiment was sight, with the only restriction being the absence of sound. The exclusion of sound aimed to emphasize the role of visual perception in the participants' interpretation of emotions through robot animations. However, it's acknowledged that involving the auditory sense could have expedited participants' recognition of emotions.

For each emotion, a specific color was associated with eye movements to enhance identification. The color selection process incorporated both a rational approach, considering the meaning of colors in art, and playful sources like cartoons, exemplified by 'Inside Out.'

Participants included a total of 20 students, comprising 11 males and 9 females. Among males, 5 were in the 18-24 age group, and 6 were in the 25-34 age group. For females, 8 were in the 18-24 age group, and 1 was in the 25-34 age group. The experiment was conducted as a *guessability* study, in which participants were asked to guess the emotion expressed by the NAO robot, choosing from a list of given emotions. The animations were presented every time in a different order, with duplicates not communicated to participants to prevent bias. Inconsistencies in participants' responses prompted interventions, where participants were asked to suggest changes or provide clarity. For more details on the experimental procedure see [2].

Results were quite encouraging: apart from Disgust (40%) and one version of Love (50%), participants quite easily guessed the emotion mimicked with gestures, see Fig. 2.

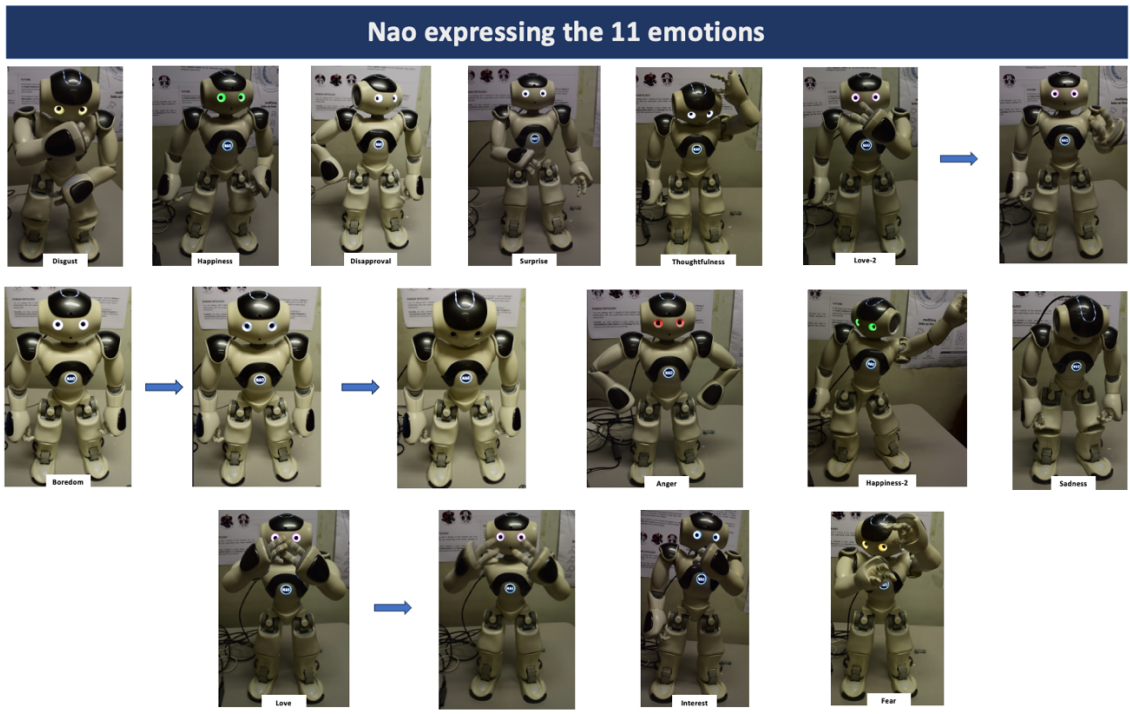

Figure 1. Nao expressing the 11 emotions, some of them is performed twice.

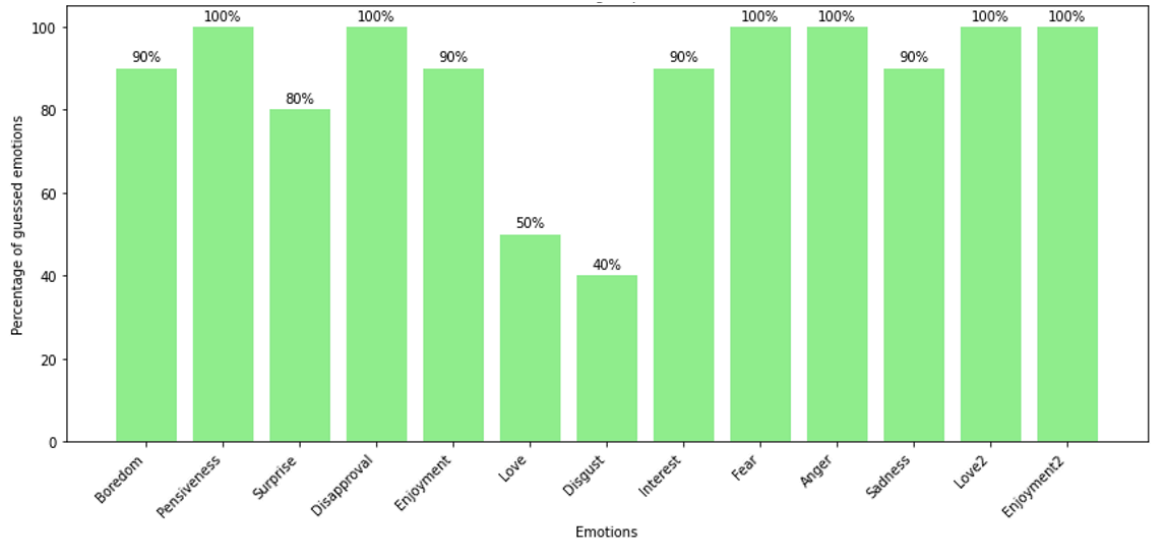

Figure 2. Experimental results of the guessability study. The percentage represents the number of participants who guessed the emotion.

## 2 DISCUSSION

The results obtained are interesting, even if we must replicate the experiment other times, and with a different sample of users. In our opinion the cultural component linked to the expression of emotions through gestures certainly influenced both us and the participants.

Italian gestures, renowned globally for their expressive qualities, often serve as an additional language that transcends linguistic barriers and have evolved into a distinct form of communication (see for instance [15]), sometimes used to express stereotypes of Italian culture [5]. While undeniably essential in conveying emotions, it would be fundamental to discern whether they hold cross-cultural significance or are more culturally specific. It would be interesting to see if these gestures are more understandable and shareable in the Mediterranean culture outside Italy than in all other countries.

Italian gestures witness the vibrant and passionate nature of the Italian people. Hand movements and facial expressions are typical of the Italian culture, and act as an extension of spoken language. This non-verbal communication emotionally enriches conversations with additional layers of meaning and nuance.

However, the meaning of Italian gestures may be less understandable in cultures where verbal communication takes precedence [11]. In societies where explicit verbal expression is used to convey emotions, elaborate gestures might be deemed redundant. In these contexts, emphasis is likely placed on word choice and tone rather than physical movements. Additionally, the understanding of Italian gestures could pose challenges in cross-cultural communication. While enhancing communication within Italian culture, these gestures might be prone to misinterpretation or overlook in different cultural settings, potentially leading to misunderstandings.

While undeniably significant within the Italian cultural context, the universal applicability and relevance of these gestures in different cultures should be further investigated, as for instance in [4].

Hence, an interesting opportunity for further exploration could be replicating our study across Mediterranean countries to ascertain whether the emotions conveyed by our 'Italian' robot are similarly recognized in settings with cultural similarities. Subsequently, a comparative analysis of the results with a subsequent study conducted in a Northern European country could yield valuable insights.

To advance this research agenda, seeking collaboration with interested researchers during this workshop could be beneficial. Collaborative efforts would involve replicating the study in different cultural contexts, allowing for a comprehensive cross-cultural examination of the reception of emotions expressed by the robot.

This proposed cross-cultural approach not only enhances the generalizability of our findings but also provides an opportunity to discern cultural influences in the interpretation of robotic gestures. It opens a dialogue for potential cultural variations in the perception of emotional cues, contributing to a wider understanding of how humanoid robots may be accepted across diverse cultural landscapes.

Initiating discussions and collaborations within the workshop setting could pave the way for a more extensive and comprehensive research attempt, making light on the cultural universality or specificity of our robot's emotional expressions.

**REFERENCES**
[1] Brunetti, D., Gena, C., & Vernero, F. (2022). Smart Interactive Technologies in the Human-Centric Factory 5.0: A Survey. *Applied Sciences*, *12*(16), 7965.
[2] Consoli, i., Gena, C., & Mattutino, C. (2024). Through NAO's Emotions: How a Robot Can Express Them Without Words, *submitted to the Socialize 2024 workshop*
[3] Ekman, P., & Friesen, W. V. (1978). Facial action coding system. *Environmental Psychology & Nonverbal Behavior*.
[4] Esposito, A., Riviello, M. T., & Bourbakis, N. (2009). Cultural specific effects on the recognition of basic emotions: A study on Italian subjects. In HCI and Usability for e-Inclusion: 5th Symposium of the Workgroup Human-Computer Interaction and Usability Engineering of the Austrian Computer Society, USAB 2009, Linz, Austria, November 9-10, 2009 Proceedings 5 (pp. 135-148). Springer Berlin Heidelberg.
[5] Gena, C., & Ardissono, L. (2001, July). On the construction of TV viewer stereotypes starting from lifestyles surveys. In *Workshop on Personalization in Future TV*.
[6] Gena, C., Mattutino, C., Perosino, G., Trainito, M., Vaudano, C., & Cellie, D. (2020, May). Design and development of a social, educational and affective robot. In *2020 IEEE Conference on Evolving and Adaptive Intelligent Systems (EAIS)* (pp. 1-8). IEEE.
[7] Gena, C., Mattutino, C., Maieli, A., Miraglio, E., Ricciardiello, G., Damiano, R., & Mazzei, A. (2021, June). Autistic Children's Mental Model of an Humanoid Robot. In *Adjunct Proceedings of the 29th ACM Conference on User Modeling, Adaptation and Personalization* (pp. 128-129).
[8] Cristina Gena, Alberto Lillo, Claudio Mattutino, and Enrico Mosca. 2022. Wolly: an affective and adaptive educational robot. In Adjunct Proceedings of the 30th ACM Conference on User Modeling, Adaptation and Personalization (UMAP '22 Adjunct). Association for Computing Machinery, New York, NY, USA, 146–150. https://doi.org/10.1145/3511047.3537684
[9] Gena, C., Manini, F., Lieto, A., Lillo, A., & Vernero, F. (2023). Can empathy affect the attribution of mental states to robots?. In Proceedings of the 25th International Conference on Multimodal Interaction, ICMI 2023, Paris, France, October 9-13, 2023 (pp. 94-103). Association for Computing Machinery.
[10] Liao, J., & Wang, H. C. (2019). Gestures as intrinsic creativity support: Understanding the usage and function of hand gestures in computer-mediated group brainstorming. *Proceedings of the ACM on Human-Computer Interaction*, *3*(GROUP), 1-16.
[11] Lutz, C., & White, G. M. (1986). The Anthropology of Emotions. *Annual Review of Anthropology*, *15*, 405–436. http://www.jstor.org/stable/2155767
[12] Macis, D., Perilli, S., & Gena, C. (2022, July). Employing Socially Assistive Robots in Elderly Care. In *Adjunct Proceedings of the 30th ACM Conference on User Modeling, Adaptation and Personalization* (pp. 130-138).
[13] Albert Mehrabian. 2017. Nonverbal communication. Routledge.
[14] Plutchik, R. (2001). The nature of emotions. *American Scientist*, *89*(4), 344–350.
[15] Poggi, I. (2002). Symbolic gestures: The case of the Italian gestionary. Gesture, 2(1), 71-98.
[16] Wobbrock, J. O., Aung, H. H., Rothrock, B., & Myers, B. A. (2005, April). Maximizing the guessability of symbolic input. In *CHI'05 extended abstracts on Human Factors in Computing Systems* (pp. 1869-1872).